\documentclass[aps,prd,twocolumn,superscriptaddress,nofootinbib,preprintnumbers]{revtex4-1}
\pdfoutput=1
\usepackage{amsmath}
\usepackage{graphicx}
\usepackage{subfigure}
\usepackage{amssymb}
\usepackage{xcolor}
\usepackage{multirow}
\usepackage{cancel}
\usepackage{color}
\usepackage{ulem}
\usepackage{listings}
\usepackage{xcolor}
\usepackage{float}
\usepackage{slashed}

\usepackage{tikz}
\usetikzlibrary{arrows,shapes}
\usetikzlibrary{trees}
\usetikzlibrary{matrix,arrows} 				
\usetikzlibrary{positioning}				
\usetikzlibrary{calc,through}				
\usetikzlibrary{decorations.pathreplacing}  
\usepackage{pgffor}							

\usetikzlibrary{decorations.pathmorphing}	
\usetikzlibrary{decorations.markings}
\tikzset{
    vector/.style={decorate, decoration={snake}, draw},
	provector/.style={decorate, decoration={snake,amplitude=2.5pt}, draw},
	antivector/.style={decorate, decoration={snake,amplitude=-2.5pt}, draw},
    fermion/.style={draw=black, postaction={decorate},
        decoration={markings,mark=at position .55 with {\arrow[draw=black]{>}}}},
    fermionbar/.style={draw=black, postaction={decorate},
        decoration={markings,mark=at position .55 with {\arrow[draw=black]{<}}}},
    fermionnoarrow/.style={draw=black},
    gluon/.style={decorate, draw=black,
        decoration={coil,amplitude=4pt, segment length=5pt}},
    scalar/.style={dashed,draw=black, postaction={decorate},
        decoration={markings,mark=at position .55 with {\arrow[draw=black]{>}}}},
    scalarbar/.style={dashed,draw=black, postaction={decorate},
        decoration={markings,mark=at position .55 with {\arrow[draw=black]{<}}}},
    scalarnoarrow/.style={dashed,draw=black},
    electron/.style={draw=black, postaction={decorate},
        decoration={markings,mark=at position .55 with {\arrow[draw=black]{>}}}},
	bigvector/.style={decorate, decoration={snake,amplitude=4pt}, draw},
}

\tikzstyle{block} = [draw, rectangle, 
    minimum height=3em, minimum width=6em]

\usepackage[colorlinks,citecolor=blue]{hyperref}
\usepackage{amsmath}
\usepackage{wrapfig}

\newcommand{\aln}[1]{\begin{align}#1\end{align}}
\newcommand{\be}{\begin{equation}}
\newcommand{\ee}{\end{equation}}
\newcommand{\beq}{\begin{equation}}
\newcommand{\eeq}{\end{equation}}
\newcommand{\bea}{\begin{eqnarray}}
\newcommand{\eea}{\end{eqnarray}}
\newcommand{\besp}{\begin{equation}\begin{split}}
\newcommand{\eesp}{\end{split}\end{equation}}

\newcommand{\Dfbd}{\mathord{\buildrel{\lower3pt\hbox{$\scriptscriptstyle\leftrightarrow$}}\over {D}_{\mu}}}

\def\0{\textbf{0}}
\def\1{\textbf{1}}
\def\2{\textbf{2}}
\def\3{\textbf{3}}
\def\4{\textbf{4}}
\def\5{\textbf{5}}
\def\6{\textbf{6}}
\def\7{\textbf{7}}
\def\8{\textbf{8}}
\def\9{\textbf{9}}

\begin{document}

\title{Late-Forming PBH: Beyond the CMB era}

\author{Philip Lu}
\email{philiplu11@gmail.com}
\affiliation{Center for Theoretical Physics, Department of Physics and Astronomy, Seoul National University, Seoul 08826, Korea}

\author{Kiyoharu Kawana}
\email{kkiyoharu@kias.re.kr}
\affiliation{School of Physics, KIAS, Seoul 02455, Korea}

\author{Alexander Kusenko}
\affiliation{Department of Physics and Astronomy, University of California, Los Angeles
Los Angeles, California, 90095-1547, USA}
\affiliation{Kavli Institute for the Physics and Mathematics of the Universe (WPI), UTIAS
The University of Tokyo, Kashiwa, Chiba 277-8583, Japan}

\begin{abstract}
The intermediate mass black hole range, $10\lesssim M_{\rm BH}^{}/M_\odot^{}\lesssim 10^{5} $, has long offered enticing possibilities for primordial black holes (PBH), with populations in this range postulated to be responsible for some of the black hole binary merger detected events as well as the existence of supermassive black holes embedded at galactic centers.  However, a prominent bound derived from PBH accretion during recombination severely restricts the mass fraction of intermediate mass PBH. We address this problem by proposing a formation scenario in which ``primordial" black holes form late in our cosmological history, beyond the CMB era, and bypassing this bound. During this crucial epoch, our population of compact objects exist as thermal balls supported by thermal pressure, which eventually cool to Fermi balls supported by degeneracy pressure and finally collapse to PBH. Our mechanism is a viable production method for both the mass gap LIGO-VIRGO-KAGRA detections and the JWST observation of an early time $z>10$ supermassive black hole. Furthermore, we present the remarkable possibility of PBH formation after the present era, which we term \textit{future} PBH. Such a population would evade most, if not all bounds on the PBH mass spectrum in the literature and open up previously unthought-of possibilities. Light future PBH could form below the Hawking evaporation threshold and convert the bulk of the matter in the Universe into radiation.
\end{abstract}

\maketitle

\section{Introduction}
\label{sec:intro}
The intermediate mass black hole (IMBH) range $\sim 10-10^{5} M_\odot$ has long been considered in connection with primordial black hole (PBH) formation, offering many unique possibilities and consequences. Although PBH as the dominant dark matter component in this mass range  is ruled out by multiple unique bounds~\cite{Lu:2020bmd,Takhistov:2021aqx,Takhistov:2021upb,Wilkinson:2001vv,2013MNRAS.428.3648I,2011ApJ...743..167B,2018PhRvD..97b3518O,Inoue:2017csr,1999ApJ...516..195C,2019PhRvL.123g1102M,Brandt:2016aco,2017PhRvL.119d1102K,Macho:2000nvd,2011MNRAS.416.2949W,2014ApJ...790..159M,Carr:2018rid, Laha:2018zav}, a subdominant population can contribute to the LIGO-Virgo-Kagra (LVK) spectrum of detected binary mergers~\cite{Nakamura:1997sm,Bird:2016dcv,Cotner:2016dhw,Cotner:2017tir,Raidal:2017mfl,Eroshenko:2016hmn,Sasaki:2016jop,Clesse:2016ajp,Cotner:2018vug,Cotner:2019ykd,Flores:2020drq,Flores:2021jas,Flores:2021tmc,Wang:2022nml}, act as seeds for the supermassive black holes (SMBH) found at the center of galaxies~\cite{Bean:2002kx,Kawasaki:2012kn,Clesse:2015wea}, and generate hypervelocity stars through three-body scattering at our galactic center~\cite{2003ApJ...599.1129Y,2006ApJ...651..392S,2019ApJ...878...17R}.

The strongest constraint in this region comes from analysis of accreting PBH, which affect recombination and distort the CMB spectrum~\cite{Ricotti:2007au}. The very restrictive limits of the CMB bound on IMBH pose problems for the PBH interpretation of PBH merger events first detected by LIGO ($f_{\rm PBH}\sim10^{-3}-10^{-4}$ and $M\sim10-100 M_\odot$)~\cite{Sasaki:2016jop} and as supermassive seeds ($f_{\rm PBH}\lesssim10^{-5}$ and $M\sim 10^{4}-10^{5} M_\odot$)~\cite{Bean:2002kx,Kawasaki:2012kn,Clesse:2015wea}. However, this bound does not apply to ``primordial" black holes formed after the CMB\footnote{Ref.~\cite{Chakraborty:2022mwu} proposed a model of PBH formation between BBN and the CMB era. Here we present an even more delayed mechanism}.

We propose a mechanism for producing these late-forming PBHs which circumvent the CMB bound. Using the asymmetric dark fermion model developed in Ref.~\cite{Hong:2020est,Kawana:2021tde,Lu:2022paj,Kawana:2022fum,Kawana:2022lba}, we consider first order phase transitions (FOPT) that produce intermediate mass compact objects existing as thermal balls~\cite{Kawana:2022lba} during recombination. These thermal balls steadily cool, eventually collapsing into Fermi balls then PBH by the present day. Although such a PBH population is prevented from being the dominant DM component by bounds from microlensing, dwarf galaxy heating, etc., the parameter space for PBH LVK binary progenitors and supermassive seeds is considerably freed. Of particular interest are the 70-80 $M_\odot$ merger event~\cite{LIGOScientific:2020iuh} and the JWST observation of an active galactic nuclei at $z\approx 10.6$~\cite{2023arXiv230207256B}.

In this paper, we develop a mechanism for producing late-forming PBH. In section~\ref{sec:thermal}, we outline the basic model and thermal history of compact objects which transform into intermediate mass PBH after recombination. In section~\ref{sec:collapse}, the conditions for the transition from thermal remnant to Fermi ball and finally to PBH are derived. In section~\ref{sec:cooling}, cooling rates and transition temperatures are calculated. In section~\ref{sec:bounds}, the relevant contraints on late-forming PBH are discussed. We summarize in Section~\ref{sec:conclusion}.

\begin{figure}[t]
\begin{center}
\includegraphics[width=.48\textwidth]{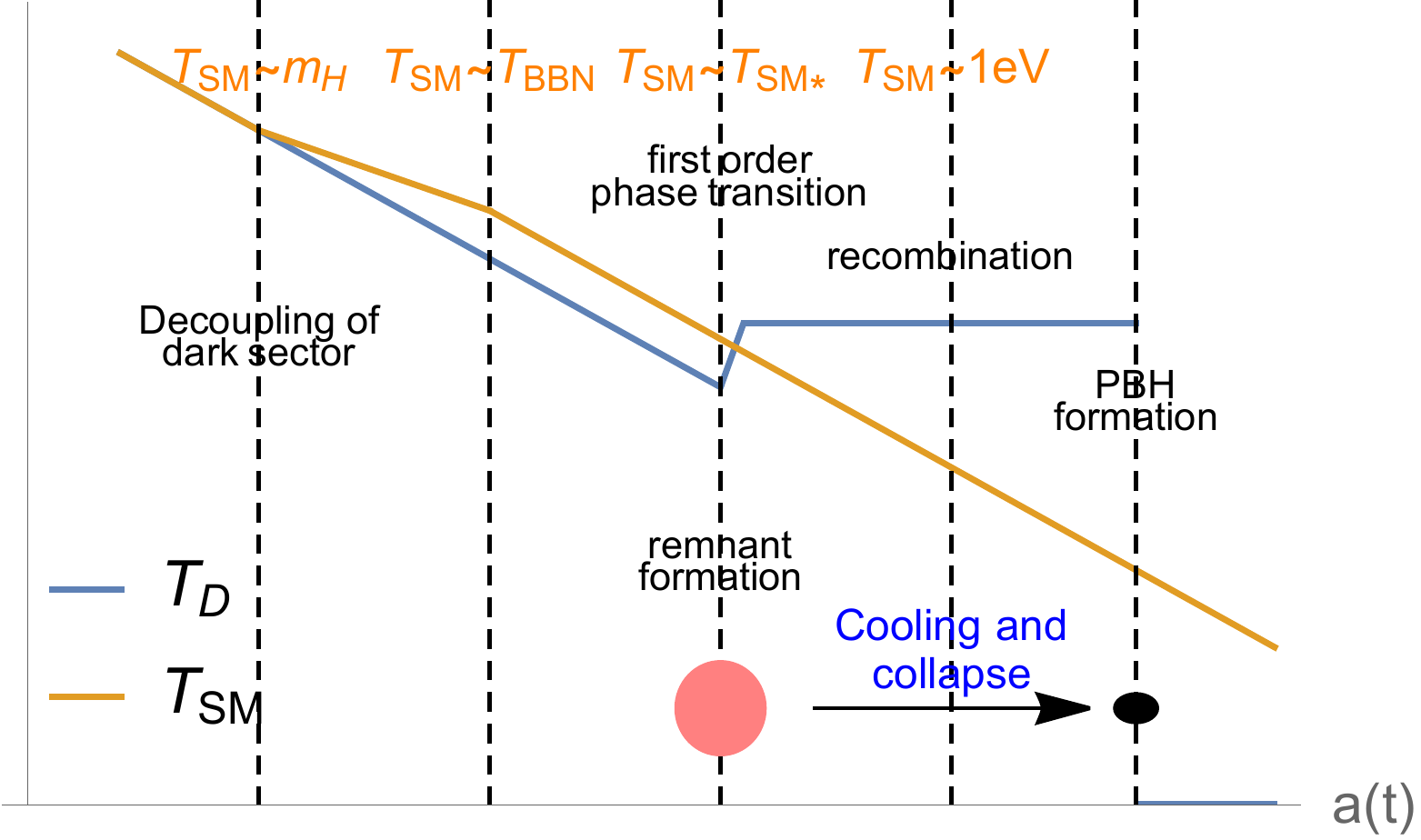}
\caption{Thermal history of our PBH formation scenario. 
The orange (blue) line corresponds to the SM (dark-sector)  temperature. 
}
\label{fig:history}
\end{center}
\end{figure}

\section{Outline}
We summarize the thermal history of PBH formation in our model and present our main points. 
The essential point in the present PBH scenario is the existence of a FOPT in the dark sector at low temperatures $T_{\rm SM}^{}\sim 1~$keV, which results in the formation of remnants and its collapse to PBHs.   
After decoupling from the SM plasma around the EW scale, the dark sector temperature $T_D^{}$ evolves independently of the SM temperature $T_{\rm SM}^{}$, and we outline the whole thermal history here to avoid confusion. 
The thermal history is schematically summarized in Fig.~\ref{fig:history}, 
and detailed calculations are presented in Section~\ref{details}.  
 
\subsection{Thermal History}
\label{sec:thermal}

We use the model of Ref.~\cite{Hong:2020est,Kawana:2021tde,Lu:2022paj,Kawana:2022fum,Kawana:2022lba} which contains a nearly dark sector with fermions $\chi$, $\Bar{\chi}$ and a scalar $\phi$ which couples to the Higgs, 
\begin{align}
\begin{split}
    \mathcal{L} =&{\cal L}_{\rm SM}^{} -\frac{1}{2}\partial_\mu \phi \partial^{\mu} \phi - \frac{\mu^2}{2}\phi^2 - \frac{\kappa}{2}\phi^2 (H^\dagger H) - U(\phi) \\ +& \Bar{\chi} i \slashed{\partial}\chi - y_\chi \phi \Bar{\chi}\chi~,
\end{split}
\end{align}
where the potential $U(\phi)$ is assumed to trigger the FOPT. 
We represent its thermal effective potential as $U_{\rm eff}^{}(\phi,T)$ in the following. 
In this paper, we do not specify a concrete shape of $U(\phi)$ (or $U_{\rm eff}^{}(\phi,T)$) in order to achieve model-independent results.  
%

%

In this nearly dark sector model, we consider the following thermal history of the Universe:  
We first assume that
both the SM and dark sector particles are in thermal equilibrium with the same temperature $T$ after inflationary reheating.\footnote{This can be achieved through the SM portal coupling $\kappa$. 
As a qualitative check, the interaction rate of $\phi\phi\leftrightarrow HH$ is 
$\Gamma\sim \kappa^2 T$, which is greater than the Hubble rate $H\sim T^2/M_{\rm Pl}$ when $\kappa\gtrsim (T/M_{\rm Pl}^{})^{1/2}$.  
Thus, as long as the reheating temperature is below $M_{\rm Pl}^{}$, $\phi$ (and $\chi$) can thermalize with the SM particles even for small values of $\kappa\ll 1$. Alternatively, a direct coupling between the inflaton and dark sector particles can result in a reheating temperature similar to that of the SM.

} 
During this epoch, the total energy density is given by
\aln{\rho(T)=\rho_{\rm SM}^{}(T)+\rho_D^{}(T)=\frac{\pi^2 g_{\rm SM}^{}}{30}T^4+\frac{\pi^2 g_{D}^{}}{30}T^4~,
}
%
where $g_{\rm SM}^{}~(g_D^{})$ denotes the effective number of degrees of freedom (d.o.f.) of the SM (dark) sector.  
The Hubble scale is given by $H^2=8\pi G\rho/3$. 

As the Universe cools down, the ordinary Electroweak (EW) phase transition occurs at around $T\sim 160$~GeV. 
In particular, the SM Higgs becomes heavy $m_H^{}=125$~GeV, and $\phi\phi\leftrightarrow HH$ processes decouple at around $T\sim m_H^{}$. 
From this moment, dark sector starts to evolve independently from the SM sector with different temperatures.     
In the following, we represent the temperature of dark (SM) sector as $T_D^{}~(T_{\rm SM}^{})$.  
%
Our PBH production mechanism proceeds as follows: 

\vspace{5mm}

\noindent 1. A FOPT occurs in the dark sector at low SM temperatures ($T_{{\rm SM}\ast} \sim 1$ keV). 
Due to the entropy conservation in each sector, their temperatures are different as 
\aln{T_{\rm SM*}^{}=\left(\frac{g_{{\rm SM},{\rm dec}}^{}}{g_{{\rm SM}*}^{}}\right)^{1/3}T_{D*}^{}~,
\label{temperature difference}
}
where $g_{{\rm SM},{\rm dec}}^{}$ ($g_{{\rm SM}*}^{}$) is the effective number of d.o.f. of the SM sector at $T_{\rm SM}^{}=m_H^{}$~$(T_{\rm SM*}^{})$. 
A moderately large expectation value of $\phi$, $v_\phi^{}(T_D^{})$, induces a large mass gap $\delta m_\phi^{}, \delta m_\chi^{} \gg T_{D*}^{}$ between the true and false vacuum, and 
dark sector particles are trapped in the false vacuum pockets of the expanding true vacuum bubbles with the trapping fraction $F_\chi^{\rm trap}\approx 1$.  
The strength parameter $\alpha_D$ of the FOPT is given by
\aln{\label{eq:alphad}
\alpha_D^{}:=\frac{1}{\rho_D^{}(T_D^{})}\left[\Delta U_{\rm eff}^{}(T_D^{})+T_D^{}\frac{\partial\Delta U_{\rm eff}^{}(T_D^{})}{\partial T}\right] \bigg|_{T_{D}^{}=T_{D*}^{}}~,
}
where $\rho_D^{}(T_D^{})$ is the energy density of dark sector and $\Delta U_{\rm eff}^{}(T_D^{})=U_{\rm eff}^{}(0,T_D^{})-U_{\rm eff}^{}(v_\phi^{}(T_D^{}),T_D^{})$. 
The other strength parameter $\beta$ which determines the duration of FOPT is given by
\aln{\frac{\beta}{H_*^{}}=-T_D^{}\frac{\partial \Gamma(T_D^{})}{\partial T_D^{}}\bigg|_{T_{D}^{}=T_{D*}^{}}~,
}
where $H_*^{}=H|_{T_{\rm SM}^{}=T_{{\rm SM}}\ast}^{}$ and $\Gamma(T_D^{})$ is the decay rate of the false vacuum.  
In this paper, we typically consider $0.1\lesssim \alpha\lesssim 1$ and $\beta/H_*^{}\sim 100$, where $\alpha$ is similarly defined as in Eq.~\eqref{eq:alphad} but relative to the SM energy density.  
\vspace{3mm}

\noindent 2. The separated false-vacuum pockets at the remnant percolation time, when the false vacuum fraction $f_0\approx 0.29$, constitute individual remnants with their typical size~\cite{Lu:2022paj}
\aln{R_*^{}\approx \frac{v_w^{}}{\beta}~.
}
where $v_w^{}$ is the bubble wall velocity. 
The corresponding asymmetrical number density of $\chi$/$\Bar{\chi}$ fermions trapped within the remnants is directly related to the eventual mass of the PBH (see Section~\ref{ssec:mass}).
The latent heat released increases the temperature of the dark sector to $T_{D,0}^{}$.

\vspace{3mm}

\noindent 3. The remnants undergo an initial shrinking phase~\cite{Kawana:2022lba} driven by the vacuum pressure, releasing latent heat and increasing the temperature of the dark sector in the false vacuum until the outward thermal pressure balances the inward vacuum pressure, forming thermal balls.\footnote{In Ref.~\cite{Kawana:2022lba}, an extremely small value of $\kappa$ was assumed to forbid the rapid annihilation $\phi\phi\rightarrow HH$. 
In the present case, $\kappa$ is still limited to small values $\lesssim 10^{-16}$ to avoid an extra contribution to the scalar mass after the EW symmetry breaking. 
} 
Since we consider a rapid phase transition with transition parameter $\beta/H_*^{} > 1$, the SM temperature stays roughly constant at $T_{\rm SM}^{}=T_{{\rm SM}\ast}
$, whereas the dark sector temperature at pressure balance increases to~\cite{Kawana:2022lba} 
\begin{equation}
    T_{D,1}^{} = \left(\frac{90 \Delta U_{\rm eff}^{}}{\pi^2 g_D^{}}\right)^{1/4}~.
    \label{balance temperature}
\end{equation}
We denote the corresponding terminal radius of the remnant in the initial shrinking as $R_1^{}$, which is given by~\cite{Kawana:2022lba}
\aln{
R_1^{}\approx \left(\frac{1+\alpha_D^{}}{4\alpha_D^{}}\right)^{1/3}R_*^{}~. 
}

\vspace{3mm}

\noindent 4. The thermal balls are unable to directly collapse to PBH, but remain at a constant temperature $T_{D,1}^{}$ 
throughout a second slow cooling phase. 
As they cool, the remnants shrink but are kept in homeostasis at $T_{D,1}^{}$ by the continuous release of latent heat.

\vspace{3mm}

\noindent 5. Eventually, the asymmetrical population of dark fermions, which remain constant in the thermal balls, dominate the thermal population which decreases in proportion to the remnant volume. 
The thermal ball then makes a transition to a Fermi ball, or a remnant supported by Fermi degeneracy pressure~\cite{Hong:2020est}. 

\vspace{3mm}

\noindent 6. After minimal further cooling, the range of the $\phi$-mediated Yukawa force 
\aln{L_\phi^{}(T_{D}^{})=m_\phi^{-1}(T_D^{})=\frac{1}{\sqrt{\mu^2+c T_D^2}}~
}
becomes long enough to cause an instability in the Fermi ball, which rapidly transitions to a PBH~\cite{Kawana:2021tde}.  
In our scenario, this final transition can happen after recombination $T_{\rm SM}^{}\sim 0.3$~eV.

\subsection{PBH Mass and Abundance
}\label{ssec:mass}

\begin{figure}[tb]
\begin{center}
\includegraphics[width=.48\textwidth]{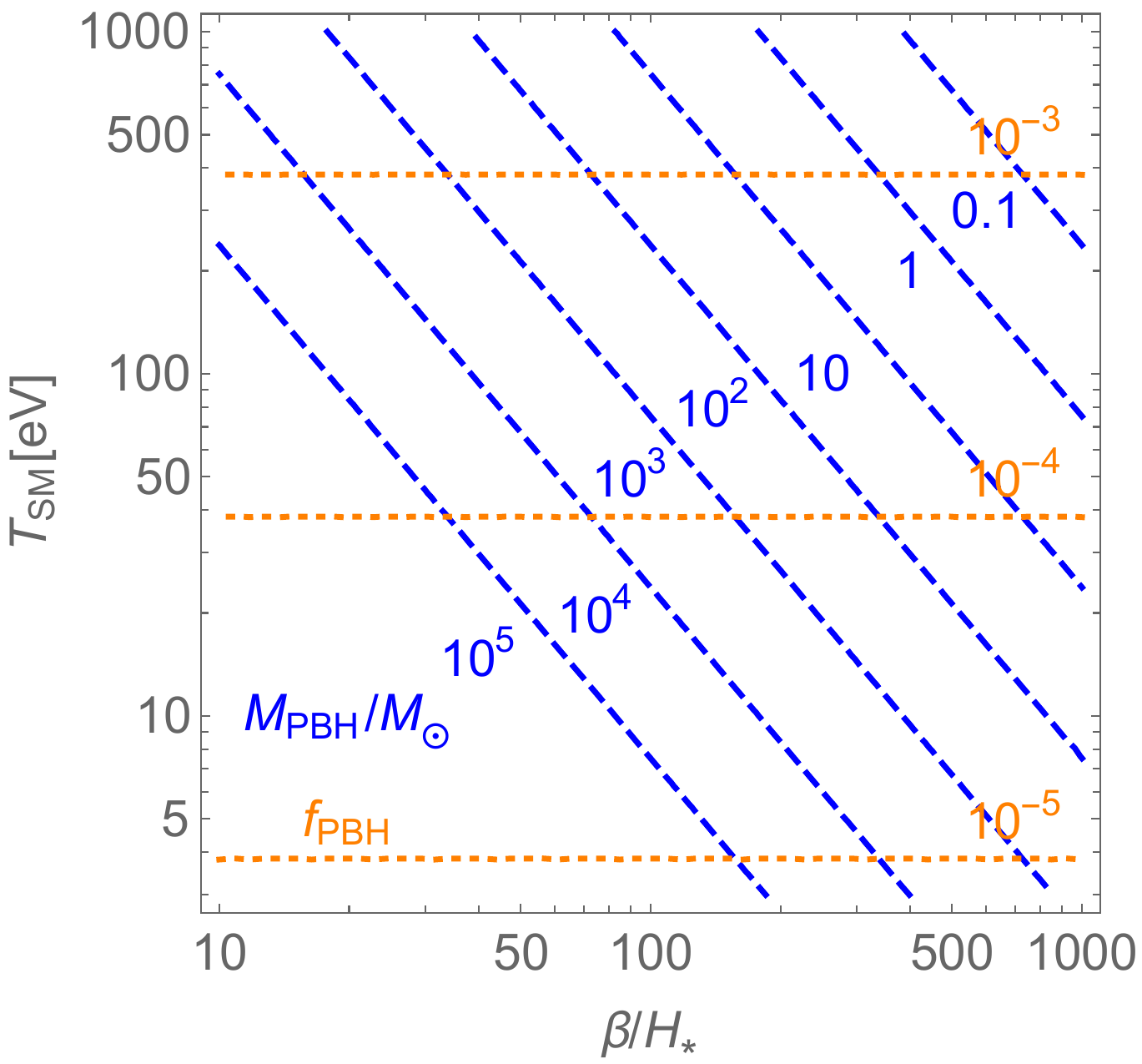}
\caption{Predicted PBH mass (blue) and abundance (orange). 
Here we take $\eta_\chi^{}=10^{-7},~v_w^{}=0.6~,\alpha_D^{}=0.1$.   
}
\label{fig:pbh}
\end{center}
\end{figure}

\noindent The average mass of the black hole resulting from Fermi ball collapse is~\cite{Kawana:2021tde}
\begin{align}
\begin{split}
\label{eq:mass}
    \overline{M}_{\rm PBH}^{} \approx & 
    10^2 M_\odot \times \alpha_D^{1/4} v_w^3 F_{\chi}^{\rm trap} 
    \left(\frac{\eta_\chi}{10^{-5}}\right)\left(\frac{\beta/H_*^{}}{100}\right)^{-3} \\ \times& \left(\frac{g_{D\ast}^{}}{4}\right)^{-1/4}\left(\frac{g_{{\rm SM}*}^{}}{g_{{\rm SM},{\rm dec}}^{}}\right)^{-2/3} 
    \left(\frac{T_{{\rm SM}\ast}^{}}{1~\rm{keV}}\right)^{-2}~. 
\end{split}
\end{align}
%
The present day PBH fraction of DM, $f_{\rm PBH}^{}:=\rho_{\rm PBH}^{}/\rho_{\rm DM}^{}$ is~\cite{Kawana:2021tde}
\begin{align}
\begin{split}
\label{eq:density}
    f_{\rm PBH}^{} \approx& 
    0.1\left(\frac{\overline{M}}{10^{2} M_\odot}\right)v_w^{-3} \left(\frac{g_{D\ast}^{}}{4}\right)^{1/2}
   \left(\frac{g_{{\rm SM}*}^{}}{g_{{\rm SM},{\rm dec}}^{}}\right)^{}  \\ \times & \left(\frac{T_{{\rm SM}\ast}}{1\textrm{ keV}}\right)^3
\left(\frac{\beta/H_*^{}}{100}\right)^3 \left(\frac{\Omega_{\rm DM}}{0.26}\right)^{-1}~.
\end{split}
\end{align}
In Fig.~\ref{fig:pbh}, we show the contours of $\overline{M}_{\rm PBH}^{}$ (blue) and $f_{\rm PBH}^{}$ (orange) in the $\beta/H_*^{}$ vs $T_{{\rm SM}\ast}^{}$ plane.
Other parameters are chosen as
\aln{\eta_\chi^{}=10^{-6}~,\quad v_w^{}=0.6~,\quad \alpha_D^{}=0.1~.
}  
There are two populations of interest which can benefit from evading the CMB bound. 
First, the hypothetical LVK population of PBH which requires black holes of mass $\lesssim 10^2 M_\odot$ and density $f_{\rm PBH}=10^{-3}-10^{-4}$. 
In our FOPT model, false vacuum remnants are only a fraction of the horizon size. 
Thus, to form intermediate mass PBH requires a much lower temperature $T_{\rm SM}^{}\lesssim $ keV (much larger horizon) when compared to the standard formation mechanism of horizon-scale perturbation collapse. 
For the proposed LVK PBH population, the typical parameter space is $\beta/H_*^{}={\cal O}(100)$ and $T_{{\rm SM}\ast}^{}={\cal O}(100~{\rm eV})$. 
%
Another interesting parameter space is for SMBH seeds, $M_{\rm PBH}^{}\gtrsim 10^4 M_\odot^{}$ and $f_{\rm PBH}^{}=10^{-5}-10^{-4}$, with lower $\beta/H_*^{}$ and $T_{{\rm SM}\ast}^{}$. 
%
%
Since the LVK population is formed at higher temperatures, a more prolonged period is necessary to delay PBH formation, whereas relatively efficient cooling can enable the SMBH seed population to form after recombination. 

\begin{table*}[t]  
\label{tab:param}
\centering 
\begin{tabular}{|c|c|c|c|c|c|c|c|}  
\hline 
 & $\overline{M}_{\rm PBH}^{}$ & $T_{{\rm SM}\ast}^{}$ & $\eta_\chi^{}$  & $\alpha$ & $\beta/H_*^{}$ & $v_w^{}$ & $f_{\rm PBH}^{}$
 \\
\hline
LVK & $30 M_\odot$ & $400$ eV & $10^{-6}$ & 0.1 & 300 & 0.6 & $10^{-3}$  \\
\hline
SMBH & $3\times 10^{4} M_\odot$ & $40$ eV & $10^{-6}$ & 0.1 & 150 &  0.6 & $10^{-4}$ \\
\hline
\end{tabular}
\caption{Benchmark parameter points for sample LVK merger and SMBH seed PBH populations. 
}
\end{table*}

In Table~\ref{tab:param}, we show two benchmark parameter values, one to represent the hypothetical LVK PBH population and the other SMBH seeds. 
%
The corresponding PBH mass spectra is plotted in Fig.~\ref{fig:latepbh} against relevant constraints using the results in 
%
Ref.~\cite{Lu:2022paj}. In particular, the yellow region corresponds to the CMB bound coming from the accretion disk emission of PBHs~\cite{Ricotti:2007au}, but this does not apply in the present PBH formation scenario. See Section~\ref{sec:bounds} for details.

\section{Detailed Thermal History}\label{details}

In the following, we will support our proposed formation scenario with detailed calculations of the cooling rate and collapse conditions. Since such late-forming PBH is a new concept, it is important to work out the viability of the model. We show that our most salient point of delayed PBH formation is well justified.

\subsection{Fermi ball Transition and PBH Formation}
\label{sec:collapse}

We first detail the process of PBH formation. After sufficient cooling, the thermal ball forms cold compact objects supported by Fermi degeneracy pressure, Fermi balls,
which then collapse into black holes.  
Here we investigate the transition of thermal balls to Fermi balls and their subsequent collapse due to the Yukawa force~\cite{Kawana:2021tde,Kawana:2022lba}. 

Let us consider the collapse condition for thermal balls, which have a lesser density and therefore are less susceptible to collapse. 
We assume a uniform density thermal ball with 
\aln{\frac{Q_{\rm therm}}{\frac{4\pi}{3}R^3}=
\frac{\zeta(3)\tilde{g}_D^{}}{\pi^2}T_D^3~,
}
where $Q_{\rm therm}^{}$ denotes the total number of particles inside a thermal ball, and $\tilde{g}_D^{}=1~(3/4)$ for each bosonic (fermionic) specie, and does not include other internal d.o.f. such as spin.    
The attractive Yukawa energy is given by~\cite{Kawana:2021tde} 
\begin{equation}
\label{eq:yukawaenergy}
    E_{\phi}^{} \approx -\frac{3y_\chi^2}{20\pi}\frac{Q_{\rm therm}^2}{R}\frac{5}{2}\left(\frac{L_\phi}{R}\right)^2= -\frac{2\tilde{g}_D^2\zeta(3)^2}{3\pi^3}y_\chi^2 R^3 T_D^6 L_\phi^2~,
\end{equation}
and the resulting pressure is
\begin{equation}
    P_\phi=\frac{\partial E_{\phi}}{\partial V} = \frac{\tilde{g}_D^2\zeta(3)^2}{2\pi^4} y_\chi^2 T_D^6 L_\phi^2~.
\end{equation}
The Fermi ball shrinks if this inward Yukawa pressure is greater than the outward thermal pressure $P_{\rm therm}^{}=\rho_D^{}/3 = (\pi^2/90)g_D^{} T_D^4$. 
If the scalar mass $m_\phi^{}(T_D^{})$ is dominated by the thermal $\chi$ term, i.e. $c=y_\chi^2/6$, then
\begin{equation}
    P_\phi^{} = \frac{\tilde{g}_D^2\zeta(3)^2}{2 \pi^4} T_D^4 < \frac{\pi^2 g_D^{}}{30} T_D^4=P_{\rm therm}^{}~
\end{equation}
for $\tilde{g}_D^{}={\cal O}(1)$. 
We see that Yukawa pressure cannot overcome thermal pressure, meaning that thermal balls are stable until they cool down sufficiently and become Fermi balls.   

The transition point from thermal to degeneracy pressure occurs around the time when the asymmetrical population exceeds the thermal population. 
The former density is given as~\cite{Kawana:2021tde}
\aln{
\begin{split}
    n_\chi^{}-\bar{n}_\chi^{}=&\left(\frac{R_1^{}}{R}\right)^3(n_\chi^{}-\bar{n}_\chi^{})\bigg|_{T_D^{}=T_{D\ast}^{}} \\ =& \left(\frac{R_*^{}}{R}\right)^3\times F_\chi^{\rm trap}\eta_\chi^{}s(T_{D\ast}^{})~,
\end{split}
} 
where $s(T_D^{})=2\pi^2g_D^{}T_D^3/45$ is the entropy density of dark sector. 
While the radius of Fermi ball $R$ steadily shrinks during the cooling phase, 
the temperature remains steady at $T_{D,1}^{}$(Eq.~(\ref{balance temperature})) due to pressure balance with the false-vacuum energy.  
The transition point (radius) is defined by the equality
\aln{
    \frac{n_\chi - n_{\Bar{\chi}}}{n_\chi+n_{\Bar{\chi}}}& =
    F_\chi^{\rm trap}\eta_\chi^{}\frac{2\pi^4}{45\zeta(3)}\left(\frac{g_D^{}}{\tilde{g}_D^{}}\right)\left(\frac{R_*^{}}{R}\right)^3\left(\frac{T_{D\ast}}{T_{D,1}}\right)^3=1
     \\
 \therefore \quad   R_{\rm tr}^{}&=0.4\times (F_\chi^{\rm trap}\eta_\chi^{})^{1/3}\left(\frac{g_D^{}}{\tilde{g}_D^{}}\right)^{1/3}\left(\frac{T_{D\ast}}{T_{D,1}}\right)R_*^{} \nonumber
 \\
 =& 7\times 10^{-3}\times {F_\chi^{\rm trap}}^{1/3}\left(\frac{\eta_\chi^{}}{10^{-5}}\right)^{1/3}\alpha_D^{-1}\left(\frac{g_D^{}}{\tilde{g}_D^{}}\right)^{1/3}R_*^{}~.
 \label{transition radius}
}
where we have used Eq.~(\ref{balance temperature}). 
This is of course comparable to the Fermi-ball radius $R_{\rm FB}^{}$~\cite{Hong:2020est,Kawana:2021tde}. 

After the transition, the Fermi ball can collapse into PBH if the Yukawa force range $L_\phi^{}$ is longer than the average separation size of $\chi~(\bar{\chi})$ particles inside the Fermi-ball~\cite{Kawana:2021tde}, which would make the remnant energetically unstable.
Explicitly, the condition for Fermi-ball collapse is given as
\begin{align}
\begin{split}
\label{eq:fermicollapse}
    & L_{\phi}(T_D^{})= m_\phi^{}(T_D^{})^{-1}=\frac{1}{\sqrt{\mu^2+cT_{D}^2}} \\ &>\frac{1}{y_\chi} \sqrt{\frac{2\pi}{3\sqrt{3}}}\left(\frac{2\pi}{3}\right)^{1/6}\frac{R_{\rm FB}}{Q_{\rm FB}^{1/3}} \approx y_{\chi}^{-1} g_{D}^{-1/4}T_{D\ast}^{-1}\alpha_D^{-1/4}~,
\end{split}
\end{align}
where $Q_{\rm FB}^{}$ is the total asymmetric charge inside the Fermi ball~\cite{Kawana:2021tde}.  

In the simple case where $\mu^2\approx0$ and
the thermal mass is dominated by the $\chi$ contribution, $c=y_\chi^2 /6$, 
%
the above condition becomes
\aln{T_{D}^{}/T_{D,1}^{}\lesssim 6g_D^{1/2}\alpha_D^{1/2}~. 
} 
Since $T_{D}^{}<T_{D,1}^{}$ after the transition from thermal balls, this result implies that Fermi balls immediately collapse into PBHs for $0.01\lesssim \alpha_D^{}$.\footnote{After the transition to Fermi ball, the finite density contribution $\sim y_\chi^{}(n_\chi^{}-n_{\bar{\chi}}^{})^{1/3}$ adds to the scalar mass, and 
%
becomes dominant over the thermal contribution as the Fermi ball shrinks. 
However, 
we have checked that PBH formation is not prevented because the inward Yukawa pressure is still stronger than the outward Fermi degeneracy pressure.  

}

%
\begin{figure}[tb]
\begin{center}
\includegraphics[width=.48\textwidth]{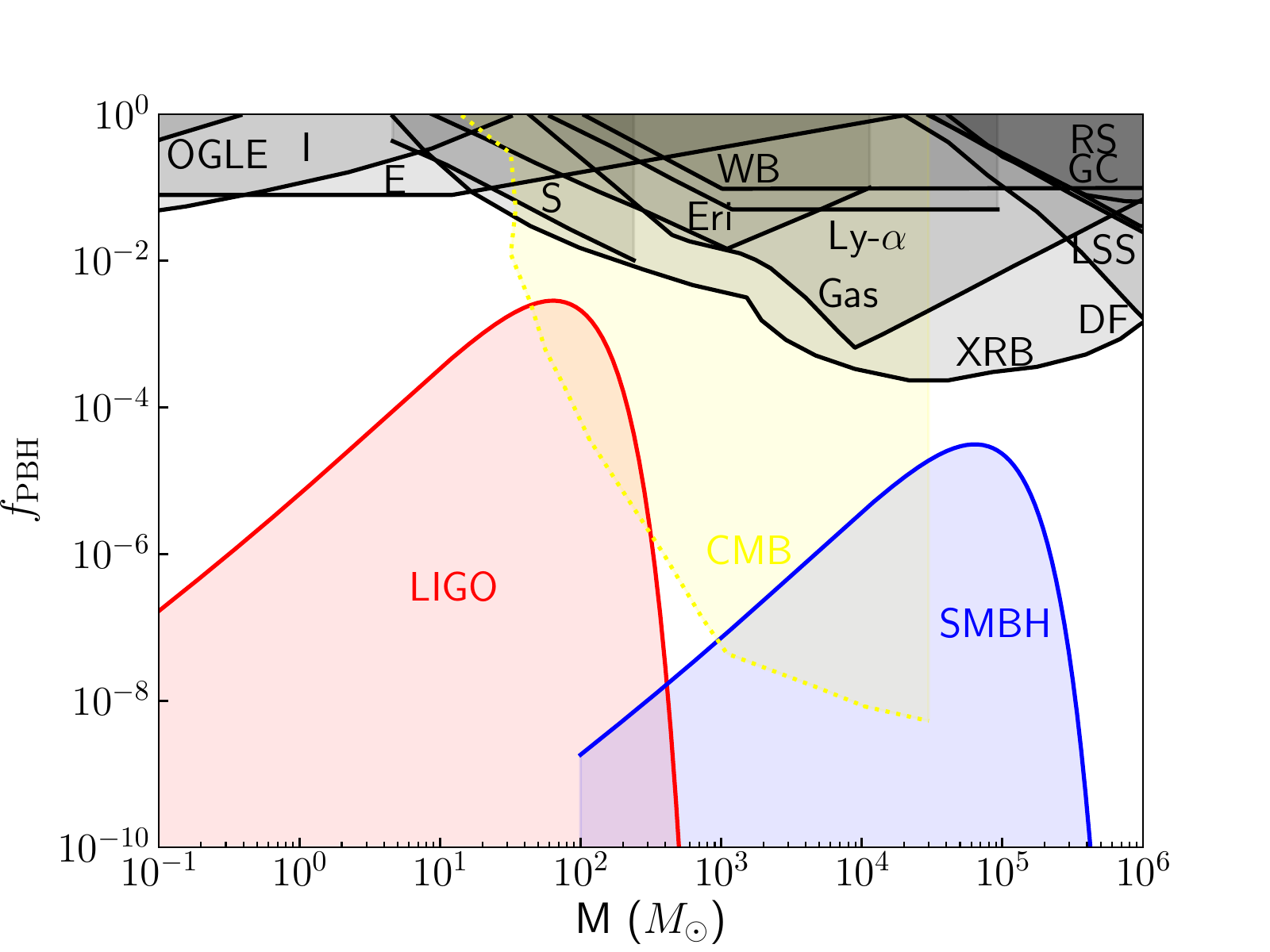}
\caption{LVK and SMBH model mass spectrums against relevant constraints in the IMBH range. The CMB bound~\cite{Ali-Haimoud:2016mbv,Poulin:2017bwe} in yellow is drawn with dashed lines to show its reduced applicability to thermal balls. Other bounds shown in black include gas heating~\cite{Lu:2020bmd,Takhistov:2021aqx} (Gas), radio sources ~\cite{Wilkinson:2001vv} (RS), globular clusters~\cite{2013MNRAS.428.3648I,2011ApJ...743..167B} (GC), Icarus~\cite{2018PhRvD..97b3518O}
(I), X-ray binaries~\cite{Inoue:2017csr}
(XRB), dynamical friction~\cite{1999ApJ...516..195C}
(DF), Lyman-$\alpha$~\cite{2019PhRvL.123g1102M}
(Ly-$\alpha$), survival of systems in Eridanus II~\cite{Brandt:2016aco} (Eri),
 Segue 1~\cite{2017PhRvL.119d1102K} (S), EROS~\cite{Macho:2000nvd}, OGLE~\cite{2011MNRAS.416.2949W} (O)
 wide binary disruption~\cite{2014ApJ...790..159M}
(WB),  large scale structure~\cite{Carr:2018rid}
(LSS), \label{fig:latepbh}
}
\end{center}
\end{figure}

\subsection{Cooling of Thermal Balls}
\label{sec:cooling}

In order to ensure that the PBH forms after recombination , $T_{\rm SM}^{}\lesssim 1~$eV, we calculate the SM temperature at which the transition from thermal ball to Fermi ball to PBH occurs. 
This temperature depends on the rate of cooling, which can proceed via surface cooling with $dE/dt =-4\pi R^2 \xi_l \rho_D^{}$ or volumetric cooling with $(4\pi/3)R^3\dot{C}$.  
Here $\xi_l$ is an efficiency parameter where $\xi_l = g_l/4 g_D$ for blackbody radiation, with $g_l$ the d.o.f. of the emitted (light) particle species, and $\dot{C}$ is the cooling rate per unit volume. 
Note that the volumetric 
cooling occurs only when the interactions among dark-sector particles are very weak (see Appendix~\ref{app:volcooling}).  
For comparison with the recombination era, it is convenient to track the evolution of the radius with the SM temperature $T_{\rm SM}^{}$.
The evolution equation is 
\begin{equation}
    \frac{dR}{dT_{\rm SM}} = \frac{dR}{dE}\frac{dE}{dT_{SM}} = \frac{(45)^{3/2} M_{\rm Pl}}{16\pi^{9/2} g_{D}^{} g_{\rm SM}^{1/2}R^2T_{D,1}^4  T_{\rm SM}^3} 
     \frac{dE}{dt}~,
\end{equation}
where we have used 
\aln{&\frac{dE}{dR}=4\pi R^2(\rho_D^{}+\Delta U_{\rm eff}^{}) =\frac{16\pi }{3}R^2\rho_D^{} \nonumber \\ & =\frac{16\pi }{3}R^2\times \frac{\pi^2 g_D^{}}{30}T_{D,1}^4~,
\\
 &\frac{dT_{\rm SM}^{}}{dt}=-HT_{\rm SM}^{}\simeq -\frac{1}{M_{\rm Pl}}\sqrt{\frac{8\pi^3 g_{\rm SM}^{}}{90}}T_{\rm SM}^3~,
}
and taken $\rho_{\rm SM} \gg \rho_D$ in the last equality. 
First, consider the case of surface cooling, $dR/dT_{\rm SM} \propto 1/T_{\rm SM}^3$, which has the solution
\aln{\label{eq:tsmtrsurf}&R(T_{\rm SM}^{})=R_1^{}-\frac{a_s}{2}\left(\frac{1}{T_{\rm SM}^2}-\frac{1}{T_{{\rm SM}*}^2}\right)~,\quad  \nonumber \\ &a_s^{}=\frac{(45)^{3/2}\xi_l^{}M_{\rm Pl}^{}}{120\pi^{3/2} g_{\rm SM}^{1/2}}~,
}
where we have assumed that $g_{\rm SM}(={\cal O}(1))$ is a constant.  
%
On the other hand, in the volumetric cooling case, $dR/dT_{SM} \propto R/T_{SM}^3$, the solution is
\begin{align}
\label{eq:tsmtrvol}
    &R(T_{\rm SM}^{}) = R_1^{}
    \exp\left[-\frac{a_v^{}}{2}\left(\frac{1}{T_{\rm SM}^2}-\frac{1}{T_{{\rm SM}*}^2}\right)\right]~,\quad \nonumber \\ &a_v^{}=\frac{(45)^{3/2}\dot{C}M_{\rm Pl}^{}}{12\pi^{7/2} g_D g_{\rm SM}^{1/2} T_1^4}~. 
\end{align}
In section~\ref{sec:collapse}, we obtained the
thermal ball to Fermi ball
transition radius $R_{\rm tr}^{}$ as a function of model parameters, i.e. Eq.~(\ref{transition radius}). 
Combining this result with specific cooling models, we will now
estimate the corresponding SM temperature of PBH formation.

\subsubsection{Blackbody Cooling}
\label{ssec:blackbody}

If the mass differences of the trapped particles between two phases are negligible, $\Delta m_i^{}\ll T_{D,1}^{}$, they can freely escape and produce blackbody radiation with $\xi_l^{}={\cal O}(0.1-1)$. 
For large couplings  $y_\chi^{}$ between $\chi$ and $\phi$, thermal equilibrium in the dark sector is maintained~\cite{Kawana:2022lba}, so that continuous emission of $\phi$ and $\chi~(\bar{\chi})$ is possible. 
%
%
Since $R_{\rm tr}^{}\ll R_1^{}$, 
the transition temperature can be qualitatively estimated by $R(T_{\rm SM}^{})\sim 0$ and it is given by
\begin{equation}
\label{eq:blackbodyttr}
    T_{\rm SM,tr} \approx \left[1+ \frac{8v_w^{}}{3\xi_l^{}}\left(\frac{\beta}{H_*^{}}\right)^{-1} \left(\frac{1+\alpha_D}{4\alpha_D}\right)^{1/3} \right]^{-1/2}\times T_{{\rm SM}\ast}^{}~.
\end{equation}
In this case, blackbody radiation rapidly cools and shrinks the remnant so that the transition to Fermi ball and PBH happens shortly after 
the phase transition. 
Thus, blackbody cooling is not conducive for late-forming PBH.

\subsubsection{Evaporation Cooling}
\label{ssec:evapcool}


On the other hand, if the dark fermions $\chi,\Bar{\chi}$ are only partially trapped $\Delta m_\chi \gtrsim 5  T_D$, then the high energy tail of the Fermi-Dirac distribution may slowly escape and cool the remnant, a scenario studied in Ref.~\cite{Kawana:2022lba}. 
If the $\phi$-mediated $\chi\chi\xrightarrow{} \chi\chi$ scattering rate is fast enough to continuously replenish the population of high energy fermions, then evaporation will be surface cooling. On the other hand, if the thermalization timescale of the remnant $\tau_{\rm therm}$ is longer than the crossing timescale $t_{\rm c}=R_1/c$, then evaporation will be volumetric cooling. Here we consider the stronger surface cooling case and leave the detailed calculation of $\tau_{\rm therm}$ and volumetric cooling for Appendix~\ref{app:volcooling}. Note that evaporation cooling from escaping $\phi$ particles is also possible and is analogous to the $\chi$,$\bar{\chi}$ cooling we consider here. 

The surface cooling factor $\xi_l^{}$ of evaporation is given in Ref.~\cite{Kawana:2022lba} as
\begin{equation}
\label{eq:surfevap}
    \xi_l = \frac{120 g_\chi}{7\pi^5 g_D^{}}\left(\frac{M_\chi}{T_{D,1}^{}}\right)^3 e^{-M_\chi /T_{D,1}^{}}~
\end{equation}
with $g_\chi^{}=4$ when there are only $\chi$ and $\bar{\chi}$ particles. 
This exponentially suppressed slow cooling rate justifies using the low transition temperature and small transition radius limit $T_{\rm SM,tr}\ll T_{{\rm SM}\ast}^{},~R_{\rm tr}\ll R_1 $ of Eq.~\eqref{eq:tsmtrsurf}, 
\begin{align}
\begin{split}
\label{eq:evapsurfsol}
    &T_{\rm SM,tr}=\left(\frac{a_s^{}}{2 R_1}\right)^{1/2}\approx 
    T_{{\rm SM}\ast}^{}v_w^{-1/2}\left(\frac{\beta/H_*^{}}{100}\right)^{1/2}\left(\frac{g_\chi}{g_D^{}}\right)^{1/2} \\&
    \times \left(\frac{1+\alpha_D^{}}{4\alpha_D^{}}\right)^{-1/6} \left(\frac{M_\chi}{T_{D,1}^{}}\right)^{3/2}e^{-\frac{M_\chi}{2T_{D,1}^{}}}~.
\end{split}
\end{align}
Then to get a transition temperature between the CMB era and the present day from an initial phase transition temperature $T_{{\rm SM}\ast}^{}\sim 1~$keV, a reduction of $10^{3}-10^{6}$ requires $20\lesssim M_{\chi}/T_{D,1}^{}\lesssim 40$. 
For even larger mass to temperature ratios, there is the interesting possibility of producing PBH from collapsing thermal balls in the future. As explained in Section~\ref{sec:bounds}, thermal balls are difficult to bound due to their less compact nature. Present day thermal balls would therefore have significantly reduced constraints across the entire mass spectrum, opening up much larger mass windows. Future populations of PBH, which evade present day bounds, may be a possible DM candidate.


\section{Bounds}
\label{sec:bounds}

All the constraints based on microlensing, accretion, etc. that are independent of cosmological history still apply to PBH forming between CMB and the present day (see Ref.~\cite{Carr:2020xqk} for a review). However, future forming PBH which exist today as slowly cooling thermal balls could evade and ameliorate most of the present day bounds on its mass spectrum.

We discuss three constraints of interest: BBN $\Delta N_{\rm eff}$, CMB bounds, and GW emission from the FOPT.
Some other bounds on PBH, such as those from considerations of large scale structure formation, may be affected although a detailed analysis is beyond the scope of this paper. While the Fermi balls are cooling, they resemble decaying dark matter, but the current bounds do not constrain a fraction of such dark matter below $\sim$0.1~\cite{Hubert:2021khy}. 

\subsection{BBN $N_{\rm eff}$}

In the models we consider, the FOPT occurs around the keV-scale so that the dark sector particles were free-floating during the BBN era. The relevant bound is the BBN $\Delta N_{\rm eff}=\rho_D/\rho_{\nu_{e}}\lesssim 0.40$~\cite{Tanabashi:2018oca} limit on the number of extra relativistic species. 
Although this limit may seem to be grossly violated because of the 2 dark sector fermions $\chi$,$\Bar{\chi}$ and scalar $\phi$. 
However, the weak coupling between the dark sector and the SM results in freeze-out at high temperatures after the EW transition. 
When the dark sector freezes out at around the EW scale,  the dark sector temperature $T_D^{}$ at BBN compared to the SM temperature $T_{\rm SM}^{}$ is $T_{D}^{}/T_{\rm SM}^{} = (10.75/106.75)^{1/3}\simeq 0.5$ (see Eq.~(\ref{temperature difference})), which corresponds to  $\Delta N_{\rm eff}^{}\simeq g_D^{}(T_{D}^{}/T_{\rm SM}^{})^4\ll 0.4$ for $g_D^{}={\cal O}(1)$.  
%
Thus, the constraint by BBN can be easily evaded.


\subsection{CMB Accretion Bound}

Here we reevaluate the Planck/CMB bound on PBH~\cite{Ricotti:2007au,2017PhRvD..95d3534A,2020PhRvR...2b3204S} for thermal ball remnants. These bounds are based largely on accretion, so the lower density and surface gravity thermal remnants are unable to convert the potential energy of accreted material into radiation as efficiently.

In Ref.~\cite{Ricotti:2007au}, the accretion disk is modeled as a thin disk with luminosity $L\sim 0.1 \dot{m}$ for high accretion rates $\dot{m}=\dot{M}/\dot{M}_{\rm Eddington}$ and as an advection dominated accretion flow $L\sim 0.011 \dot{m}^2$ for low accretion rates. 
In each of these accretion models, the luminosity is roughly proportional to the surface gravitational potential, which is calculated at the innermost stable circular orbit radius for PBH, $GM/3R_s = 1/6$ for Schwarzchild black holes.
In contrast, the minimum radius for thermal balls is much larger and consequently, the surface gravity is much lower. For simplicity, we assume a uniformly distributed thermal ball so that $M=(4\pi R^3 /3) \rho_D^{}$ and $\rho_D^{} = g_D^{} (\pi^2/30)T_D^4$ so that the ratio of surface potentials between thermal remnants and PBH is
\begin{equation}
    \frac{(GM/R)_{\rm therm}}{(GM/R)_{\rm PBH}} = G\frac{4\pi^3}{45} g_D R^2 T_{D,1}^4~.
\end{equation}
The surface potential of the thermal ball is highest immediately after formation and decreases during the slow shrinking phase. The ratio of the potentials at this time is
\begin{align}
&\frac{(GM/R)_{\rm therm}}{(GM/R)_{\rm PBH}}\bigg |_{R=R_1^{}} \approx 2\times10^{-5} \left(\frac{g_D}{g_{{\rm SM}\ast}}\right)
v_w^2 \nonumber \\ & \times \left(\frac{\beta/H_*^{}}{100}\right)^{-2} \frac{(1+\alpha_D)^2}{\alpha_D} ~.
\end{align}
For typical parameters, the thermal ball surface potential even at its maximum is much weaker than for PBH of similar mass. Thus, the accretion based CMB bounds should be weaker by the same factor, i.e., $f_{\rm therm}< 10^{6} f_{\rm PBH}\sim 10^{-3}$ at its peak assuming the most stringent conditions. Since the thermal balls are formed around $\sim 1$ keV, by the recombination era of $\sim 1$ eV, the radius will have decreased further, weakening the bound. Additional second-order effects such as the lower temperature of the thermal ball accretion disk result in proportionally less ionizing X-rays emitted. PBH accretion following recombination may still induce spectral distortions into the CMB. However, for PBH formation at $z\lesssim \mathcal{O}(100)$ the gas density should be sufficiently low. Therefore, the CMB bound on PBHs does not significantly apply to pre-collapse thermal balls and is superseded by the other non-cosmological PBH bounds in the IMBH mass range.

\subsection{Gravitational Wave Bounds}
\begin{figure}[t]
\begin{center}
\includegraphics[width=.48\textwidth]{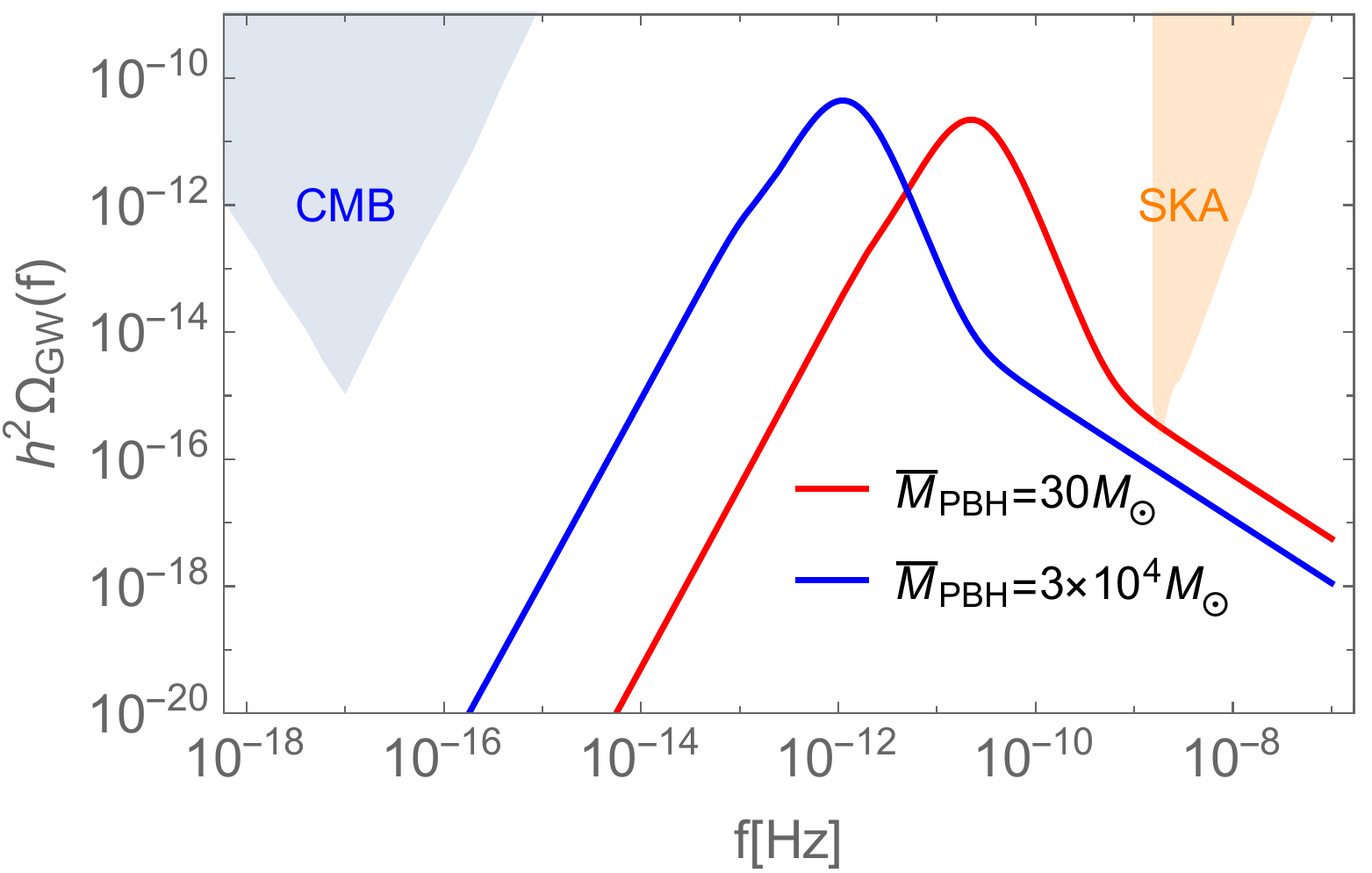}
\caption{GW spectra for the two benchmark parameter points presented in Table~\ref{tab:param}. 
The red (blue) contour corresponds to $\overline{M}_{\rm PBH}=30M_\odot^{}~(3\times 10^4 M_\odot^{})$. 
}
\label{fig:GW}
\end{center}
\end{figure}

Another phenomenological signature of FOPT is the production of stochastic GWs~\cite{Caprini:2015zlo,Caprini:2019egz,Schmitz:2020syl}. 
%
In general, there are three
sources of the stochastic GWs produced during a FOPT: bubble collisions, sound waves, and magnetohydrodynamic (MHD) turbulence in the plasma. 
%
%
Although the GW strength from these sources have different parameter dependences, 
their peak frequencies are qualitatively the same, 
\aln{
f_{\rm peak}^{}={\cal O}(10^{-2}-10^{-3}{\rm mHz})\times \left(\frac{\beta}{H_*^{}}\right)\left(\frac{T_{{\rm SM}\ast}^{}}{100~{\rm GeV}}\right)~,
}
which means that the typical peak frequencies in our PBH scenario are $f_{\rm peak}^{}={\cal O}(10^{-11}-10^{-12}{\rm Hz})$ for $\beta/H_*^{}=100$ because $T_{{\rm SM}*}^{}\lesssim 1~$keV .  
%
These frequencies are too low even for pulsar timing array experiments such as the International Pulsar Timing Array (IPTA)~\cite{Lentati:2015qwp,Shannon:2015ect,NANOGRAV:2018hou,Aggarwal:2018mgp} and the Square Kilometre Array (SKA)~\cite{Zhao:2013bba} to detect. 
Note that there also exists the bound by CMB observations~\cite{Namikawa:2019tax} in the low frequency region, $f\lesssim 10^{-15}$~Hz, which is well below the frequencies in our model.  
Thus, there are no bounds from GW signals in the present scenario.    
We explicitly show the GW spectra in Fig.~\ref{fig:GW} for the
parameters of the two benchmark models presented in Table~\ref{tab:param},
%
using the the numerical fitting functions in Ref.~\cite{Caprini:2015zlo}.

\section{Conclusion}
\label{sec:conclusion}

In this paper, we have presented a viable method for producing late-forming PBH which exist in thermal ball form until late in our cosmological history. 
The primary benefit of this novel production scenario is to avoid constraints on the PBH mass spectrums which rely on cosmology, most notably the CMB bound in the intermediate mass range. 
The bypassing of this constraint allows otherwise restricted ``primordial" populations to generate some of the (higher mass) merger events detected by LVK and to seed the puzzlingly supermassive black holes found at the center of galaxies and clusters. 
To support our claims, we have presented detailed calculations of collapse conditions and cooling rates to show that such prolonged formation is indeed possible. 
 Building upon previous works in this series, Refs.~\cite{Kawana:2021tde,Lu:2022paj,Kawana:2022lba}, we have selected sets of model parameters that generate extended mass distributions of both LVK and SMBH seed populations, showing our results in Fig.~\ref{fig:latepbh}. We provide a valid formation mechanism for the LVK mass gap black holes as well as the JWST $z\approx10.6$ SMBH that is unencumbered by CMB bounds.

In addition to PBH forming between the CMB era and the present day, we propose an even more exotic possibility of thermal ball remnants collapsing in the future to Fermi balls and PBH. 
This ghost population of PBH could circumvent an even larger subset of PBH bounds, so that intermediate mass PBH comprising the whole of dark matter could be possible at later epochs. 
We therefore term these compact objects future PBH. 
Future PBH could open up many exciting paths in our cosmological evolution. 
Since light $M<10^{17}~g$ thermal balls would not be affected Hawking evaporation, they would be exempt from limits in that mass range. 
Future PBH could form around or below the mass threshold of $10^{15}~g$, rapidly emitting Hawking radiation, converting the bulk of the matter density into radiation. 
We leave exploring the possibility of a second radiation era from future light PBH to future work.

\section*{Acknowledgements} 
We would like to thank Ke-Pan Xie, TaeHun Kim, and Volodymyr Takhistov, Marco Flores, Edoardo Vitagliano, and Kazunori Kohri for useful discussions.
The work of PL is supported by Grant Korea NRF-2019R1C1C1010050.
K.K. would like to thank Yukawa Institute for Theoretical Physics, Kyoto University for the support and the hospitality during his stay by the long term visiting program.
A.K. was supported by the U.S. Department of Energy (DOE) Grant No. DE-SC0009937, by the Simons Foundation Fellowship, by the World Premier International Research Center Initiative (WPI), MEXT, Japan, by Japan Society for the Promotion of Science (JSPS) KAKENHI grant No. JP20H05853, and by the UC Southern California Hub with funding from the UC National Laboratories division of the University of California Office of the President.

\appendix

\section{Volumetric Evaporation Cooling}
\label{app:volcooling}

Here we calculate the thermalization time $\tau_{therm}$ which will relate the volumetric cooling rate $\dot{C}$ to the parameters of the FOPT and enable a quantitative condition for the dominance of surface cooling at large couplings (more interactions and faster thermalization) to volumetric cooling at small couplings.

From Ref.~\cite{Kawana:2022lba}, the scattering cross section of dark sector particles can be approximated by the dominant process of $\chi \chi \xrightarrow{} \chi \chi$ scattering,
\begin{equation}
\label{eq:dsigmadt}
    \frac{d\sigma}{dt} = \frac{y_\chi^4}{16\pi s^2}~,
\end{equation}
where $s= (p_1+p_2)^2 =4E_{\rm cm}^2$ and $t=(p_1-p_3)^2$ are the usual Mandelstam variables. The energy change $\delta E$ of the colliding particle 1 after its collision with particle 2 is
\begin{equation}
    \delta E = \frac{(E_1-E_2)(\cos \phi - 1)}{2} + \sqrt{E_1 E_2 (1+\cos\theta)/2}\sin \phi~,
\end{equation}
where $\theta$ is the angle between the particle momenta before the collision in the thermal ball frame and $\phi$ is the (post-)scattering angle in the CM frame. By symmetry, both angles are spherically distributed. We use a thermal Fermi-Dirac distribution for both particles 1 and 2 as a function of $\epsilon=p/T$
\begin{equation}
\label{eq:fermidirac}
    f(\epsilon)=\frac{1}{e^\epsilon+1}~.
\end{equation}
Using the relations
\begin{equation}
     \frac{d\sigma}{d \phi}= \frac{d\sigma}{dt} \frac{dt}{d \phi} = \frac{y_\chi^4 \sin \phi}{64\pi E_1 E_2 (1-\cos\theta)}~,
\end{equation}
and $v_{\rm rel} \simeq 2c$, rate of energy transfer of a particle 1 with energy $E_1 = \epsilon_1 T$ is
\begin{align}
\begin{split}
    \frac{\delta E_1}{\delta t} =& \frac{T^2 y_\chi^4}{512 \pi^3 \epsilon_1}\int_{0}^\infty d \epsilon_2 \frac{\epsilon_2}{e^{\epsilon_2}+1}\int_{0}^{\pi} d\theta \frac{\sin\theta}{(1-\cos\theta)} \\ \times &\left(\int_{0}^{\pi} -\int_{\pi}^{2\pi}\right)d\phi \sin^2 \phi \left(\frac{(\epsilon_1-\epsilon_2)(\cos\phi-1)}{2} \right. \\ + & \left. \sqrt{\frac{\epsilon_1 \epsilon_2 (1+\cos\theta)}{2}}\sin\phi\right) = 2.582\times10^{-4} \frac{T^2 y_\chi^4}{\sqrt{\epsilon_1}}~.
\end{split}
\end{align}

To find the thermalization timescale, average across the incident particle energy distribution
\begin{equation}
    \tau_{\rm therm} = \frac{2}{3\zeta(3)}\int_{0}^{\infty} d\epsilon_1 \frac{\epsilon_1^2}{e{\epsilon_1}+1} \frac{E_1}{\delta E_1/\delta t} = \frac{2.402\times10^{4}}{T y_\chi^4}~.
\end{equation}
Comparing this to the crossing timescale 
\begin{equation}
    t_c = R_1/c= \frac{1.22 v_w}{\beta}\left(\frac{1+\alpha_D}{4\alpha_D}\right)^{1/3}~,
\end{equation}
at temperature $T=T_{D,1}$ (Eq.~\eqref{balance temperature}), the cooling will be surface dominated if $t_c > \tau_{\rm therm}$,
\begin{align}
\begin{split}
    y_\chi >& 9.50\times10^{-6} \left(\frac{T_{\rm SM\ast}}{1\textrm{ keV}}\right)^{1/4} \left(\frac{\beta}{H}\right)^{1/24} \\ \times & g_{\rm SM\ast}^{1/6}g_{\rm SM,dec}^{1/12}\left(\frac{4\alpha_D}{1+\alpha_D}\right)^{1/12}~,
\end{split}
\end{align}
and volume cooling dominated if the opposite inequality is satisfied.

The volumetric cooling rate $\dot{C}$ from evaporation cooling is~\cite{Kawana:2022lba}
\begin{equation}
\label{eq:volevap}
    \dot{C}=\frac{g_\chi T M_\chi^3}{2\pi^2 \tau_{\rm therm}}e^{-M_\chi/T} = \frac{g_\chi y_\chi^4 T^2 M_\chi^3}{4.74\times10^{5}}e^{-M_\chi/T} ~,
\end{equation}
where we have substituted the previous result for $\tau_{\rm therm}$.

As with the surface cooling case, the volumetric cooling rate is exponentially suppressed. Therefore the use of the small transition temperature limit of Eq.~\eqref{eq:tsmtrvol} is justified. The transition temperature in this regime is then
\begin{align}
\begin{split}
\label{eq:evapvolsol}
    &T_{\rm SM,tr} = \left(\frac{a_v}{2 \ln(R_1/R_{tr})}\right)^{1/2} = 1.7 \textrm{ eV} \left(\frac{y_\chi}{10^{-6}}\right)^2 \left(\frac{T_1}{\textrm{keV}}\right)^{1/2} \\
    & \times \left(\ln\left[\frac{R_1}{R_{\rm tr}}\right]\right)^{-1/2} \left(\frac{M_\chi}{T_1}\right)^{3/2} e^{-M_\chi/2T_1}~.
\end{split}
\end{align}
For this boundary value of $y_\chi$, transitioning between the CMB and the present era requires $10\lesssim M_\chi/T_1 \lesssim 25$.

\section{SM Portal Cooling}
\label{app:SMcooling}

The thermal ball remnant can cool by SM particle production via the Higgs portal. Since SM particles are not trapped by the bubble wall, they can easily escape. For most of the cooling, $T_{SM} \ll T_{D,1}$, as the remnant is heated up during the initial collapse, the SM plasma within the thermal ball is negligible. Thus, the energy in the produced particles freely disburses into the surrounding low temperature plasma.

The dominant production channel at the low temperatures of interest, $T\lesssim 1 \textrm{ keV}$, is $\phi\phi\xrightarrow{} \gamma\gamma$ with cross section
\begin{align}
\begin{split}
\label{eq:gammacross}
    \sigma \simeq& \frac{\alpha^2 e^2 (\kappa v)^2}{1152\pi^3 \sin \theta_W M_W^2}\frac{s}{(s-M_H^2)^2}~, \textrm{ for } s\ll m_t^2 \\
    \approx & 2.717\times10^{-9} \frac{\kappa^2 s}{M_H^4}~.
\end{split}
\end{align}
Integrating over the incident and target particle Fermi-Dirac momentum distribution, and incoming angle $\theta$, the cooling rate is calculated to be
\begin{align}
\begin{split}
\label{eq:smcool}
    \dot{C} =& \frac{T^3}{2\pi^2} \int d\epsilon_1 \frac{\epsilon_1^2}{e^{\epsilon_1}+1} \frac{T^3}{2\pi^2}\int d\epsilon_2 \frac{\epsilon_2^2}{e^{\epsilon_2}+1} (\epsilon_1+\epsilon_2)T \\ \times & \int_{0}^{\pi} d\theta \frac{\sin \theta}{2} v_{\rm rel} \frac{d\sigma}{ds}\frac{ds}{d\theta} =  5.808\times10^{-9} \frac{\kappa^2 T^9}{M_H^4}~,
\end{split}
\end{align}
where we have used $v_{\rm rel}=2$, $s=2\epsilon_1\epsilon_2(1-\cos\theta)$, and the factor $(\epsilon_1+\epsilon_2)T$ is the energy loss to the SM. The cooling rate is minimal, so that the low transition temperature limit of Eq.~\eqref{eq:tsmtrvol} gives
\begin{align}
\begin{split}
    T_{\rm SM,tr} \simeq & \left(\frac{a_v}{2 \ln(R_1/R_{\rm tr})}\right)^{1/2} = 2.8\times10^{-6} \textrm{ eV} \kappa \\ \times & \left(\frac{T_1}{\textrm{keV}}\right)^{5/2} \left(\ln\left[\frac{R_1}{R_{\rm tr}}\right]\right)^{-1/2}~.
\end{split}
\end{align}
For the transition temperature to be between CMB $T_{SM}\sim 1$ eV and the present day $T_{SM}\sim 2\times10^{-4}$ eV, the thermal ball temperature should be in the range of $6\textrm{ keV} \kappa^{-2/5} \lesssim T_{D,1} \lesssim 170\textrm{ keV}\kappa^{-2/5}$. SM portal cooling is therefore not conducive to forming intermediate mass PBH by the present day unless there is extreme reheating of the dark sector particles from the FOPT temperature of $\sim 1\textrm{ keV}$ to $\sim 1 \textrm{ TeV}$. However, the extreme slow cooling of the thermal ball remnants is ideal for the scenario of \textit{future} PBH as discussed in Section~\ref{ssec:evapcool}.

\bibliographystyle{apsrev}
\bibliography{references}

\end{document}